\documentstyle[twoside,fleqn,epsfig,espcrc2]{article}

\def\lsim{\raise0.3ex\hbox{$\;<$\kern-0.75em\raise-1.1ex\hbox{$\sim\;$}}}
\def\gsim{\raise0.3ex\hbox{$\;>$\kern-0.75em\raise-1.1ex\hbox{$\sim\;$}}}

\newcommand{\AmS}{{\protect\the\textfont2
  A\kern-.1667em\lower.5ex\hbox{M}\kern-.125emS}}

\title{$b \rightarrow s \gamma$ and CP violation in the MSSM}

\author{ O. Vives\address{SISSA -- ISAS, Via Beirut 4, I-34013, Trieste, 
Italy and \\ INFN, Sezione di Trieste, Trieste, Italy.}}

\begin{document}

\begin{abstract}
In this work we study possible new contributions to $\varepsilon_{K}$ and
$\varepsilon_{B}$ in the MSSM with large supersymmetric phases.
We show that, in the CMSSM, the constrains coming from the experimental 
measure of the $b \rightarrow s \gamma$ decay imply that these contributions
are too small to be detected in CP violation experiments with the 
available sensitivity.

\end{abstract}

\maketitle
The minimal supersymmetric extension of the SM (MSSM) contains new observable
phases which can cause deviations from the predictions of the SM in CP 
violation experiments. In fact, in the so--called Constrained 
Minimal Supersymmetric Standard Model (CMSSM) with strict universality at 
the GUT scale there are two new phases present. These phases can be chosen 
to be the phases of the $\mu$ parameter ($\varphi_{\mu}$) and the trilinear 
soft coupling, ($\varphi_{A_{0}}$).

It is well--known that for most of the CMSSM parameter space the 
experimental bounds on the electric dipole moments of the electron and 
neutron constrain $\varphi_{A_0,\mu}$ to be at most ${\cal{O}}(10^{-2})$.
So, these new supersymmetric phases have been taken to vanish 
exactly in most studies of CMSSM.
However, in the last few years the possibility of having non-zero SUSY phases
has again attracted a great deal of attention. Several new mechanisms have 
been proposed to suppress EDMs below the experimental bounds while allowing 
SUSY phases ${\cal{O}}(1)$ \cite{cancel}. 

In this work we are going to study new effects on CP--violation observables 
in the CMSSM with large supersymmetric phases.

\section{Flavor change in the CMSSM}

The CMSSM is completely defined at the electroweak scale in terms of 
$\tan \beta$, the scalar mass $m_0^2$, the gaugino mass $m_{1/2}$, the 
trilinear coupling $A_0$ and the two phases, $\varphi_A$, $\varphi_\mu$ 
when we require radiative symmetry breaking \cite{bertolini}.
Even in this simple model with strict universality, due to the existence of 
two Yukawa matrices non simultaneously diagonalizable, 
some flavor mixing leaks through RGE into the sfermion mass matrices. 
In fact, in the SCKM basis, any off-diagonal entry in 
the sfermion mass matrices at $M_W$ will be necessarily proportional to a 
product of Yukawa couplings. In the up (down) squark mass 
matrix the up (down) Yukawas will mainly contribute to diagonal entries 
while off--diagonal entries will be due to the down (up) Yukawa matrix.
This means, for instance, that in this model the off-diagonality in the 
$m^{(d)\,2}_{LL}$ matrix will roughly be $c \cdot Y_u Y_u^\dagger$. Where $c$ 
is a proportionality factor that, in order of magnitude, is
roughly $ c \simeq 1/(4 \pi)^2 \log (M_{Gut}/M_W) \simeq 0.20$
as expected from the loop factor and the running from $M_{GUT}$ to $M_W$.
On the other hand, it is also clear that these flavor changing entries in the 
down squark mass matrix will be very stable with $\tan \beta$, as for 
$\tan \beta \gsim 1$ the up Yukawa matrix is approximately the same for any
$\tan \beta$. For the same 
reasons, the $\tan \beta$ dependence is very strong in the up squark mass 
matrix because the down Yukawa matrix grows linearly with $\tan \beta$ for 
large $\tan \beta$.
All these are well--known facts in the different studies of FCNC processes
in the framework of the CMSSM \cite{bertolini} and imply that Flavor mixing is
still dominantly given by the usual CKM mixing matrix in W-boson, charged
Higgs and chargino vertices. 

In this analysis we are specially interested 
on CP violating observables, and then we must also consider the presence
of observable phases in the sfermion mass matrices.      
In the following we take $\delta_{CKM}=0$ to isolate 
pure effects of the new supersymmetric phases on CP violating observables.
Then, before RGE evolution the susy phases ($\varphi_A$, $\varphi_\mu$) are 
confined to the left-right part of the sfermion mass matrix while both the 
left-left, $m^2_{Q}$, and right-right, $m^2_{D\, ,U}$, are real diagonal 
matrices. However this is not strictly true anymore at $M_W$: $\varphi_A$ 
leaks into the off-diagonal elements of these hermitian matrices through 
RGE evolution. 
From the explicit RGE in the MSSM \cite{bertolini}, it is clear that this 
phase only enters the $(m_Q^2)_{i j}$ evolution through the combinations 
$(A_U A_U^\dagger)_{i j}$ or $(A_D A_D^\dagger)_{i j}$. At $M_{GUT}$ these 
matrices have a common phase and so the combination $(A A^\dagger)$ is exactly
real. So, to the extent that the $A$ matrices keep a 
uniform phase during RGE evolution no phase will leak into the $m_{LL}^2$ 
matrices. However, this is not the case and different elements of the $A$ 
matrices are renormalized differently. At $M_W$ the general form of this 
matrix in terms of the initial conditions is,
\begin{eqnarray}
m_{Q}^{2}(M_{Z})=\eta^{(m)}_{Q} m_{0}^{2}+\eta^{(A)}_{Q} A_{0}^{2}+
\eta^{(g)}_{Q} m_{1/2}^{2}\\+ \Big(\eta^{(g A)}_{Q} e^{i\varphi_{A}} + 
\eta^{(g A)\,T}_{Q} e^{- i\varphi_{A}}\Big) A_{0} m_{1/2} 
\nonumber
\end{eqnarray}
where the coefficients $\eta$ are $3\times 3$ matrices with real numerical 
entries. In this expression we can see that the presence of imaginary parts
will be linked to the non-symmetric part of the $\eta^{(g A)}_{Q}$ matrices.
We have explicitly checked that these non-symmetric parts of $m_Q^2$ are 
present only in one part per $2-3 \times 10^3$ \cite{paper}. This situation 
was also found by Bertolini and Vissani in the CMSSM without new phases for 
the leakage of $\delta_{CKM}$ \cite{vissani}. So, we conclude that in the 
processes we will consider we can take the $m^{(u)}_{LL}$ and $m^{(d)}_{LL}$
as real to a very good approximation.   

\section{Indirect CP violation in the CMSSM}

In the SM neutral meson mixing arises at one loop through the well--known
$W$--box. However, in the CMSSM there are new contributions to $\Delta F=2$
processes coming from  boxes mediated by supersymmetric particles. These are,
charged Higgs boxes ($H^{\pm}$), chargino boxes ($\chi^{\pm}$) and 
gluino-neutralino boxes ($\tilde{g}$, $\chi^{0}$). The amount of the indirect
CP violation in the neutral meson ${\cal{M}}$ system is measured by 
the well--known $\varepsilon_{{\cal{M}}}$ parameter, and it depends on the
 matrix elements of the $\Delta F=2$ Hamiltonian, 
${\cal{H}}_{eff}^{\Delta F=2}$,
\begin{eqnarray}
\label{DF=2}
{\cal{H}}_{eff}^{\Delta F=2}=-\frac{G_{F}^{2} M_{W}^{2}}{(2 \pi)^{2}}
(V_{td}^{*} V_{tq})^{2}( C_{1}(\mu) Q_{1}(\mu) \nonumber \\
+C_{2}(\mu) Q_{2}(\mu) +C_3(\mu) Q_3(\mu))
\end{eqnarray}
where the operators are  $Q_{1}=(\bar{d}^{\alpha}_{L}\gamma^{\mu}
q^{\alpha}_{L})^2$, $Q_{2}=(\bar{d}^{\alpha}_{L}q^{\alpha}_{R})^2$ and 
$Q_{3}=\bar{d}^{\alpha}_{L} q^{\beta}_{R} \cdot \bar{d}^{\beta}_{L}
q^{\alpha}_{R} $. Here, $q=s , b$ for the $K$ and $B$--systems respectively 
and $\alpha, \beta$ are color indices. 
These are the only three operators present in the limit of vanishing $m_d$. 

The three operators in Eq (\ref{DF=2}) are very different with respect to the
flavor mixing in the sfermion mass matrices. 
The operator $Q_1$ preserves chirality along the fermionic line while the 
operators $Q_2$ and $Q_3$ change chirality. This means, for 
instance, that gluino contributions to $C_1$ will not need a chirality 
change in the sfermion propagator and hence will involve mainly the 
$m^{(d)}_{LL}$ submatrix, real to a very good approximation. The operators 
$Q_2$ and $Q_3$ always involve a change in the chirality of the external 
quarks and consequently also a change of the chirality of the associated 
squarks or gauginos.
This implies the presence of the new supersymmetric phases.

In first place we will consider the Wilson coefficient $C_1$. All the 
different supersymmetric boxes contribute to this operator.
Both the usual SM $W$--box and the charged Higgs box do contribute to $C_1$, 
however, in the absence of CKM phase, they can only contribute to the mass 
difference $\Delta M_{\cal{M}}$, but never to the imaginary part in 
$\varepsilon_{\cal{M}}$.
In the gluino contribution the source of flavor mixing is not directly the 
usual CKM matrix, but it is the presence of off--diagonal elements in the 
sfermion mass matrices discussed in the previous section. Working in the 
SCKM basis all gluino vertices are flavor diagonal and real. This means that 
in the MI approximation we need a complex mass insertion in one of the 
sfermion lines. We have seen in the previous section that these mass 
insertions are real in one part per $2 \times 10^{-3}$. The values for the 
real and imaginary parts of the mass insertions required to saturate 
$\Delta M_{\cal{M}}$ and $\varepsilon_K$ are, 
$\sqrt{|Re (\delta^d_{12})^2_{LL}|} < 4 \cdot 10^{-2}$ and
$|(\delta^d_{1 2})_{L L}| \sin (2\phi_{L L}) < 3 \cdot 10^{-3}$.   
Taking into account the relation found in the previous section
between real and imaginary parts, respecting the bound on the real part
implies that no sizeable contributions to $\varepsilon_K$ can be found. 
The situation in $B^0$--$\bar{B}^0$ mixing is completely analogous. 

The chargino can also contribute to $C_{1}$. In this case flavor mixing 
comes explicitly from the CKM mixing matrix and flavor mixing in the
sfermion mass matrices plays only a secondary role.
In the approximation of no intergenerational mixing in the sfermion mass 
matrices we have \cite{branco},
\begin{eqnarray}
\label{chWC}
C_1^\chi (\mu_{0}) = \frac{1}{4} \sum_{i,j=1}^{2} \sum_{k, l=1}^{6} 
\sum_{\alpha \beta} \frac{V_{\alpha d}^{*} V_{\alpha q}V_{\beta d}^{*} 
V_{\beta q}}{(V_{td}^{*} V_{tq})^2} \\
\left( G^{(\alpha,k)i} G^{(\alpha,k)j*} G^{(\beta,l)i*} 
G^{(\beta,l)j}  Y_1(z_{k}, z_{l}, s_i, s_j)\right) \nonumber
\end{eqnarray}
where $z_k = M_{\tilde{u}_k}^2/M_W^2$, $s_i =M_{\tilde{\chi}_i}^2/M_W^2$, and 
$G^{(\alpha,k)i}$ represent the coupling of chargino and squark $k$ to 
left--handed down quarks. All these couplings and the loop function 
$Y_1(a,b,c,d)$ can be found in \cite{branco}. 
From Eq.(\ref{chWC}), taking into account that $ Y_1(a,b,c,d)$ is symmetric 
under the exchange of any pair of arguments we can easily see that this
contribution is exactly real. 
We have explicitly checked that the presence of intergenerational mixing in 
the sfermion mass matrices does not modify this fact at an observable level.
Imaginary parts appear at least two orders of magnitude below the 
corresponding real parts. Hence we will not consider them here, more details 
will be given in a complete paper \cite{paper}. 

Now we analyze the contributions to the $C_2$ and $C_3$ Wilson coefficients.
The charged Higgs box contributes only to $Q_2$, but once again
the absence of phases prevents it to contribute to $\varepsilon_{\cal M}$.
Gluino boxes contribute both to $Q_2$ and $Q_3$, however flavor change 
will be given in this case by a left-right mass insertion that in the CMSSM is 
always proportional to the mass of the right handed squark.  
This mass insertions have phases ${\cal O}(1)$ but the mass suppression will 
not be compensated in any case by a large value of $\tan \beta$.
This means that gluino contributions will always be smaller than the 
corresponding chargino ones. 

The most important contribution will usually be the chargino box. Before the 
inclusion of QCD effects it contributes solely to the coefficient $C_3$. 
At first approximation CKM produces directly all the necessary flavor change, 
then we have, 
\begin{eqnarray}
\label{chWCR}
C_3^\chi (M_W) = \sum_{i,j=1}^{2} \sum_{k,l =1}^{2} 
[ F_s(3,k,3,l,i,j)-\\
2 F_s(3,k,1,l,i,j)+F_s(1,k,1,l,i,j)]\nonumber
\end{eqnarray}
with the functions $F_s(\alpha,k,\beta,l,i,j)=H^{q(\alpha,k)i}G^{(\alpha,k)j*} 
G^{(\beta,l)i*} H^{q(\beta,l)j} Y_2(z_{\alpha k}, z_{\beta l}, s_i, s_j)$ 
and $H^{q(\alpha,k)i}$ the coupling of chargino and squark 
to the right--handed down quark $q$ \cite{branco}. We have 
used CKM unitarity and degeneracy of the first two generation squarks.
In this case, due to 
the differences between the $H$ and $G$ couplings this contribution is always 
complex in the presence of susy phases.
Recently we showed in a short letter \cite{fully}, that this contribution 
could be relevant for $\varepsilon_{\cal M}$ in the large $\tan \beta$ regime.
However, in that work we did not take into account the additional constrains
coming from $b \rightarrow s \gamma$ decay.
In the next sections we will analyze the relation of $\varepsilon_{\cal M}$
with this decay, and the constrains imposed by the experimental measure.

\section{$b\rightarrow s \gamma$ decay}

The decay $b \rightarrow s \gamma$ has already been extensively studied in the 
context of the CMSSM with vanishing susy phases
\cite{bsgama}. Being the branching ratio a CP conserving 
observable, the presence of new phases will not modify the main features found 
in \cite{bsgama}.
However we will see that the experimental measure will also have a large 
impact on the imaginary parts of the decay amplitude.
This decay is described by the following $\Delta F=1$ effective Hamiltonian 
\begin{eqnarray}
{\cal H}_{eff}^{\Delta F=1}=-\frac{4 G_{F}}{\sqrt{2}} V_{t s}^{*}V_{t b}
\sum_{i=2,7,8} {\cal{C}}_{i} {\cal{O}}_{i}
\end{eqnarray}
where the different operators involved are ${\cal{O}}_{7}=e m_{b}/(4 \pi)^{2}\bar{s}_{L}
\sigma^{\mu\nu}F_{\mu \nu}b_{R}$ that contributes directly to the decay and 
${\cal{O}}_{2}=\bar{s}_{L} \gamma_{\mu} c_{L} \bar{c}_{L} \gamma^{\mu} b_{L}$
and  ${\cal{O}}_{8}= g_{s} m_{b}/(4 \pi)^{2} \bar{s}_{L}\sigma^{\mu\nu}
G_{\mu\nu}b_{R}$, that contribute through RGE running.
Here ${\cal{C}}_{2}(M_W)=1$, and the photonic and gluonic penguins receive 
contributions from $W^\pm$ boson, charged Higgs $H^{\pm}$, chargino 
$\chi^{\pm}$, neutralino $\chi^{0}$, and gluino $\tilde{g}$ loops.

Among these contributions, the $W$ penguin is exactly the same as in the SM
and does not depend on any supersymmetric parameters. Similarly, the charged 
Higgs penguins only depend on $m_{h^\pm}$ from the new susy parameters. 
The expressions for ${\cal{C}}^{W, H}_{7}(M_W)$ and 
${\cal{C}}^{W, H}_{8}(M_W)$ can be found for instance in 
\cite{bertolini,bsgama}. 

Then, we have also diagrams mediated by neutralinos and gluinos. Flavor change 
in this diagrams is due to the off--diagonality in the sdown mass matrix.
However, this flavor change is always smaller than direct CKM mixing and 
being a left--right off--diagonal mass insertion is suppressed by $m_b$. 
Indeed, smallness of the neutralino and gluino contributions has
already been established in \cite{bsgama} where it was 
shown that, in the CMSSM, such contributions are roughly one order of
magnitude smaller than the chargino contribution. 

The most important supersymmetric contribution will be in a large part of the 
parameter space the chargino contribution. This is due to the fact that in
this diagram, the chirality change can be made both through a chargino mass 
insertion in the loop or an external leg mass insertion proportional to $m_b$.
We are then mainly interested in the chargino mass insertion. In terms of the 
chargino--quark--squark couplings used in the previous section, these 
contributions are,
\begin{eqnarray}
{\cal{C}}_{7,8}^{\chi^{\pm}}(M_W)=\sum_{k=1}^{2}\sum_{i=1}^{2}
\frac{m_{\chi^{i}}}{m_{b}} ( H^{b (3, k) i}{G^{*}}^{(3, k) i}\nonumber \\
F_{R}^{7,8}(z_{3 k},s_{i})
- H^{b(1, k) i}{G^{*}}^{(1, k) i}
F_{R}^{7,8}(z_{1 k},s_{i}) ) 
\label{char}
\end{eqnarray} 
where the loop functions are defined by 
$F_{R}^7(x,y)=[F_{1}(y/x)+ e_U F_{2}(y/x)]/x$ and 
$F_{R}^8(x,y)=  F_{4}(y/x)/x$ with the functions $F_{i}(x)$ in 
\cite{bertolini}. Once more we use CKM unitarity and degeneracy of 
the first two generations of squarks. We can see in this expression
that the enhancement due to $m_{\chi^{i}}/m_{b}$ is partially compensated by 
the presence of the $b$ yukawa coupling in $H^{b (\alpha, k) i}$, 
\cite{bertolini}. However this contribution will still be too large for large 
values of $\tan \beta$.

\section{ $b \rightarrow s \gamma$ and $\varepsilon_{\cal M}$ correlated
analysis}

If we compare Eqs. (\ref{chWCR}) and (\ref{char}) we can see that, except
the presence of different loop functions and possibly different Yukawa 
couplings of the down quarks, both chargino contributions to the $C_3$ and
${\cal C}_7$ Wilson coefficients are deeply related.
In fact, the couplings $H^{q (\alpha, k) i}$ only depend on the down quark
$q$ through its Yukawa coupling $h_q = m_q/(\sqrt{2} M_W \cos \beta)$.
If we make a rough approximation and we take all the loop functions in the sum
equal for both $C_3$ and ${\cal C}_7$ we would obtain that $C_3$ is exactly 
the square of ${\cal C}_7$ times $m_q^2/M_W^2$. Naturally, this is 
not at all a good approximation, but we can expect that the order of
magnitude of $C_3$ is determined by the allowed values of ${\cal C}_7$. 
\begin{figure}[htb]
\vspace{9pt}
\epsfig{file=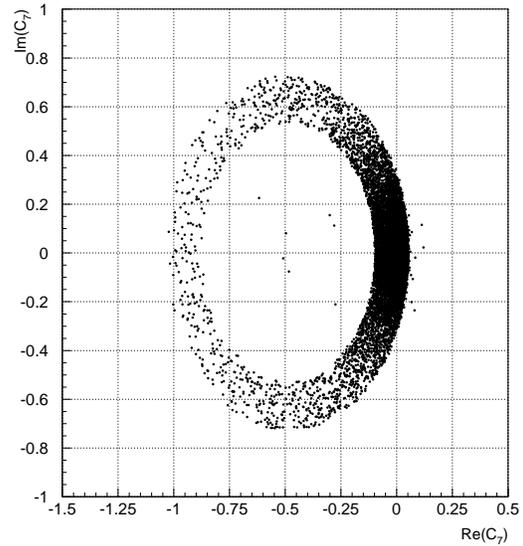,width= \linewidth}
\caption{Experimental constraints on the Wilson Coefficient ${\cal C}_7$ }
\label{fig:largenenough}
\end{figure}
\begin{figure}[htb]
\epsfig{file=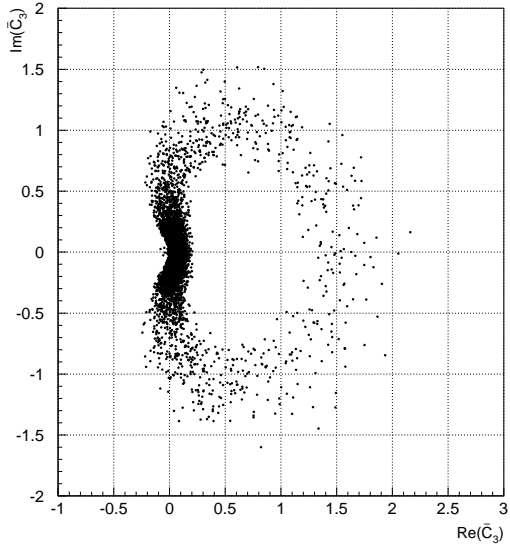,width= \linewidth}
\caption{Allowed values for the re--scaled WC $\bar{C}_3$}
\label{fig:toosmall}
\end{figure}
Following \cite{kagan-neubert}, it is possible to 
constrain in a model--independent way new physics contributions to the 
Wilson coefficients at $M_W$. In terms of these Wilson coefficients
$\mbox{BR}(B\rightarrow X_{s} \gamma)$ is,
\begin{eqnarray} 
\label{constrain}
\mbox{BR}(B\rightarrow X_{s} \gamma)\simeq 1.17 + 0.38 |\xi_{7}|^{2} + 
0.015 |\xi_{8}|^{2}+ \\ 1.39 Re[\xi_{7}]+ 0.156 Re[\xi_{8}]
+0.083 Re[\xi_{7}\xi_{8}^{*}] \nonumber
\end{eqnarray}
where $\xi_{a}={\cal{C}}_{a}(M_W)/{\cal{C}}_{a}^{W^{\pm}}(M_W)$, and the 
values of the different coefficients are taken from \cite{kagan-neubert}.
Then, with the experimental measure, 
$\mbox{BR}(B\rightarrow X_{s} \gamma)= (3.14 \pm 0.48) \times 10^{-4}$, we can 
constrain the allowed values of the complex variables $\xi_{7}$ and $\xi_{8}$. 
In fact, we can already see from 
Eq. (\ref{constrain}), that in the approximation $\xi_7 \approx \xi_8$, this 
is simply the equation of an ellipse in the $Re[\xi_7]$--$Im[\xi_7]$ plane.
In the case of supersymmetry the new physics contribution to 
$\xi_{7}$ and $\xi_{8}$ will be mainly due to the chargino contributions as 
we have discussed in the previous section. The allowed values of 
$\xi_7$ constrain then directly the chargino contributions to 
${\cal C}_7(M_W)$ and indirectly the values of 
$C_3(M_W)$. 
  
In figure 1 we show a scatter plot of the allowed values of $Re(\xi_{7})$
versus $Im(\xi_{7})$ in the CMSSM for a fixed value of $\tan \beta =40$
with the constrains from Eq. (\ref{constrain}). The fact that now $\xi_7$ and
$\xi_8$ are independent does not modify strongly the shape of the plot.
This value of $\tan \beta$ could give rise to observable CP violation 
\cite{fully}.
However the shape of the plot is clearly independent of $\tan \beta$, only
the number of allowed points and its location in the allowed area depend on 
the value of $\tan \beta$ considered.
Figure 2 shows the allowed values for the re--scaled Wilson coefficient
$\bar{C}_3(M_W)= M^2_W/M_q^2 C_3(M_W)$ corresponding to the same parameter
space points shown in figure 1.
As expected the allowed values for $\bar{C}_3$ are close to the square 
of the values of ${\cal C}_7$ in figure 1 slightly scaled by different values 
of the loop functions.
We can immediately translate this result to a constrain on the size of the 
chargino contributions to $\varepsilon_{\cal M}$. 
\begin{eqnarray}
\label{epscoef}
\varepsilon_{\cal M}=\frac{G_F^2 M_W^2 F_{\cal M}^2 M_{\cal M} \eta_3}
{4 \pi^2 \sqrt{2}\ \Delta M_{\cal M}}
\frac{(V_{td} V_{tq})^2}{24}\frac{M^2_{\cal M}}{m_q^2} Im[C_3]
\end{eqnarray}
So, for $\varepsilon_K$ we have $C_3= m_s^2/M_W^2 \bar{C}_3$ and this implies 
that $\varepsilon_K^\chi \lsim 0.5 \times 10^{-7}$. Then in this simple model,
even with large susy phases, $\varepsilon_K$ will be mainly given by the 
usual SM box. Similarly $\varepsilon_B^\chi \approx 0.4 \times 10^{-3} 
Im[\bar{C}_3]$ and hence out of reach in the forecoming B factories.

\end{document}